\documentclass[conference,a4paper]{IEEEtran}

\newcommand{\remove}[1]{}

\newtheorem{definition}{Definition}
\newtheorem{lemma}{Lemma}

\newtheorem{theorem}{Theorem}

\usepackage[cmex10]{amsmath}
\usepackage{amssymb}
\usepackage{mathrsfs}
\usepackage{graphicx}
\usepackage{color}
\usepackage{cite}
\usepackage{multirow}
\usepackage{framed}

\newcommand{\bP}{{\mathbb P}}

\newcommand{\cR}{{\cal R}}
\newcommand{\cC}{{\cal C}}
\newcommand{\cM}{{\cal M}}

\newcommand{\cH}{{\cal H}}

\newcommand{\cS}{{\cal S}}
\newcommand{\cL}{{\cal L}}

\newcommand{\bF}{{\mathbb F}}
\newcommand{\SD}{{\bf SD}}

\newcommand{\enc}{\mathsf{AWTPenc}}
\newcommand{\dec}{\mathsf{AWTPdec}}

\newcommand{\ADV}{{\bf Adv}}

\newcommand{\pl}{\mathsf{Poly}}
\newcommand{\rorw}{{ (\rho_r, \rho_w) }}

\newcommand{\edwtp}{{$(\epsilon, \delta)$-AWTP}}

\ifCLASSINFOpdf
\else
\fi

\begin{document} 

\title{ Efficient Codes for Adversarial Wiretap Channels}

\author{Pengwei~Wang and Reihaneh~Safavi-Naini \\
Department
of Computer Science, University of Calgary,
Canada\\
 e-mail: [pengwwan,  rei]@ucalgary.ca 
}
\remove{
\markboth{Journal of \LaTeX\ Class Files,~Vol.~11, No.~4, December~2012}
}

\maketitle

\begin{abstract}
In \cite{PS13} we proposed a $(\rho_r, \rho_w)$-{\em adversarial wiretap channel model (AWTP)} in which 
the  adversary can adaptively choose to see a fraction  $\rho_r$ of the codeword sent over the channel,
 and modify a fraction  $\rho_w$ of the  codeword 
by adding arbitrary noise values to them. 
\remove{
We derived the secrecy capacity of these channels and gave the construction of a capacity achieving code 
family. 
}
In this paper we give the first efficient  construction 
of a capacity achieving code family 
 that provides  perfect secrecy for this channel.
\end{abstract}

%
\IEEEpeerreviewmaketitle

%
%
%
%

\section{Introduction} \label{intro}

In Wyner's  wiretap model \cite{W75} channel   noise in the channel is
used as a resource for the system designer 
to provide (asymptotic) perfect secrecy  against a computationally unbounded adversary
without the need for a shared key. 
In this model,
a  sender and a receiver communicate over a noisy channel referred as the main channel,
and their communication is eavesdropped 
by an adversary  through a second noisy channel,
 referred to as the adversary channel. 
 The goal is to provide (asymptotic) perfect reliable communication from sender to receiver with 
(asymptotic) perfect secrecy against the adversary.  In this model adversary is passive and 
 obstruction of its view by noise is probabilistic.

Recently a number of models \cite{ALCP09,BS13,MBL09} that include a  stronger adversary that can modify communication
have been introduced. These models primarily use arbitrarily varying channel approach and assume 
eavesdropper and jammer (who modifies communication) do not communicate.
We introduced \cite{PS13} an adversarial model for wiretap channel in which the adversary can adaptively
choose a fraction of the communicated codeword to see and  a fraction to modify.
The modification of each component is by adding (algebraic) an arbitrary value (adversary's choice) to the component.
The adversary's choice of observation and tampering components is unrestricted, as long as the total number of observation and tampering symbols
 are within specific limits. 
An Adversary Wiretap Channel (AWTP) is specific by two  parameters $(\rho_r, \rho_w)$ and is denoted by
$(\rho_r, \rho_w)$-AWTP channel.   
An $(\epsilon, \delta)$-AWTP code guarantees that the information leaked about the message (measured using statistical distance) and the probability of decoding failure are upper bounded by $\epsilon$ and 
 $\delta$, respectively.  The {\em information rate of   a code $C$} is 
 $R(C)= \frac{\log |\cM|}{ N\log |\Sigma|}$ where $N$ is the length of the code and $\cM$ is the message space.
  The code provides {\em perfect secrecy if $\epsilon=0$.}

\remove{For a  codeword of  length $N$, the adversary read and tampering sets are $S_r$ with length $|S_r|=\rho_rN$
 choose a set of codeword $S_r$ with length $|S_r|=\rho_rN$ to read and a set $S_w$ with $|S_w|=\rho_wN$ to write. The adversary wiretap code is defined to transmit information securely and reliability over adversary wiretap channel.}
We derived an upper bound on the rate of codes for  $(\rho_r, \rho_w)$- AWTP channels as
$R(\mathbb{C})\leq 1-\rho_r-\rho_w+2\epsilon\log_{|\Sigma|} \frac{1}{\epsilon}$, and code family with perfect secrecy is $R(\mathbb{C})\leq 1-\rho_r-\rho_w$. An explicit and {\em inefficient} construction of AWTP code is also given in \cite{PS13}.

\subsection{Our Result}\label{sec_intro_codefamily}

We give an {\em efficient} construction of a 
 code family $\mathbb{C}=\{C^N ; N\in \mathbb{Z}\}$ in which every  code $C^N$ of  length $N$,  provides 
perfect secrecy for a $\rorw$-AWTP channel.
The construction  uses three building blocks: an Algebraic Manipulate Detection Code (AMD code) \cite{CDFP08},  a Subspace Evasive Sets. (SES) \cite{DL12}, and
a Folded Reed-Solomon code (FRS code) \cite{Gur11}.
AMD code detects algebraic manipulation assuming the adversary is oblivious and does not have access to the codeword.
SES are subsets with the property that their intersection with any subset of certain dimension is bounded.
FRS code is a special class of Reed-Solomon code that achieve list decoding capacity, and have efficient encoding and decoding.
Encoding of a message uses the three building blocks in order: the message is encoded using AMD code, then 
using a SES and finally an FRS code.
In decoding, first the FRS decoder  outputs a list of possible codewords. 
This list for the decoding algorithm in \cite{Gur11}, is a function of $N$, the code length.
Using the intersection algorithm of SES the list can be pruned to a shorter list which is
independent of the code length.
The final step is to use the AMD code  to find the correct message. The decoder always outputs  the correct
message.
We prove with appropriate choice of parameters,
each code in the family  is perfectly secure, satisfies the upper bound on rate for $(\rho_r, \rho_w)$-AWTP channels with equality and so is capacity achieving,
 and finally the probability of decoding error reduces exponentially in $N$. 

\subsection{Related Work}

Wiretap channels have been an active area of research for a number of years with excellent progress on extending the model and strengthening security against passive adversary \cite{BTV12,BB11,CK78,LH78,MV11,MBL09,OW84}.
More recently active adversary for these channels have been considered \cite{ALCP09,BS13,MBL09,PS13}.
The  active adversary in \cite{BS13,MBL09} is modeled using 
arbitrarily varying channels, and is assumed that there is no communication between the eavesdropper and the wiretapper.
 In \cite{ALCP09} the wiretap II model is extended to active adversary.  The adversary however
 is  restricted to flip the codeword components that they have chosen to read. 
 \remove{ \cite{13}\cite{MBL09} study the arbitrarily varying wiretap channel with the transition probability specified by probability $\bP[y, z|x, s]$ where $s$ is the state of channel. The jammer can corrupt the transmission of main channel by affecting the states $s$. The previous work of active adversary in wiretap setting fall short on one or more of the following fall short on one or more of the following, (i) considering sufficiently strong adversary’s capability, (ii) using a strong definition of security, (iii) deriving an expression or a tight upper-bound for secrecy capacity, and (iv) providing an efficient or explicit construction. 
}
In \cite{SW13} we proposed a model for adversarial channel called {\em limited view adversarial channel (LVAC)},
 which is the same as the adversarial channel considered here.
The goal of communication however was reliability only. $\rorw$-AWTP channels have the same adversary power as LVAC channel, 
but the goal of communication is reliability and privacy both.

\smallskip

\noindent 
{\em Paper orgnization:}
In section \ref{sec_awtp}, we recall  the model and capacity results for $(\rorw)$-AWTP channels.
In section \ref{sec_construction}, we give our construction and conclude the paper in section \ref{sec_conclude}.


\section{Model and Definitions}\label{sec_awtp}
We consider the following scenario. 
Alice (Sender $\cS$) wants to a send messages $m \in {\cal M}$ securely and reliably to Bob (Receiver $\cR$), over  a communication channel that is partially controlled by  Eve (Adversary). 
 Let $[N]=\{1,\cdots, N\}$. $S_r= \{i_1,\cdots, i_{\rho_rN}\} \subseteq [N]$ and $S_w= \{j_1,\cdots, j_{\rho_wN}\} \subseteq [N]$ denote two subsets of the $N$ coordinates. For a vector $x$, $\mathsf{SUPP}(x)$ denotes the set of coordinates where  $x_i$ is non-zero. 
Let $\Sigma$ denote the code alphabet, with an underlying group operation.

\begin{definition}\cite{PS13}\label{def_awtpchannel}
A {\em$(\rho_r, \rho_w)$-Adversarial Wiretap channel} ($(\rho_r, \rho_w)$-AWTP channel), is an adversarially corrupted 
 communication channel between Alice and Bob such that it is (partially) controlled by an adversary Eve, with two capabilities: Reading and Writing.
In {\em Reading} (or {\em Eavesdropping }),   Eve selects a subset $S_r\subseteq [N]$ of size at most $\rho_rN$
 and sees the components of the sent codeword $c$ on $S_r$. Eve's view of the codeword is the set of all read components:
$\mathsf{View}_{\cal A}(\enc(m, r_{\cal S}), r_{\cal A})=\{c_{i_1},\cdots, c_{i_{\rho_rN}}\}.
$
In {\em Writing} (or {\em Jamming}), Eve  chooses
 a subset $S_w\subseteq [N]$ of size  at most $\rho_wN$
and adds an error vector $e$ to  $c$, where  the addition is component-wise and
over $\Sigma$.
We require  $\mathsf{SUPP}(e) = S_w$.
 The corrupted components of $c$ are
$\{y_{j_1},\cdots, y_{j_{\rho_wN}}\}$  and $y_{j_\ell}=c_{j_\ell}+e_{j_\ell}$. The error $e$ is generated according to the Eve's 
best strategy to make  Bob's decoder fail.
\end{definition}
	
The adversary is {\em adaptive} and selects  components of $c$ 
for reading and writing, one by one and at each step using 
its knowledge of the codeword at that time.

Alice and Bob will use an {\em Adversarial Wiretap Code} to provide security and reliability for communication over Adversary wiretap channel.
\begin{definition}\label{def_awtpcode}\cite{PS13}
An  {\em $({\cal M}, N, \Sigma,\epsilon, \delta)$-AWTP Code} ($(\epsilon, \delta)$-AWTP code for short) for a $(\rho_r, \rho_w)$-AWTP channel consists of a randomized encoding $\enc : {\cal M}\times {\cal U} \rightarrow \cC$, from the message space $\cM$ to a code $\cC$,
and a deterministic decoding algorithm $\dec :  \Sigma^N \rightarrow \{ {\cal M}\; \cup \perp\}$, such that 
$\dec(\enc(m, r_{\cal S}))=m$ for all $m\in \cM$.
The code guarantees secrecy and reliability as defined below.\\
i) {\em Secrecy:}  For any two messages $m_1, m_2 \in \cM$,
we have 
\[
\begin{split}
&\ADV^{\mathsf{ds}}(\enc, \mathsf{View}_{\cal A})\stackrel{\triangle}=\\
&\qquad\qquad\max_{m_0, m_1}\SD(\mathsf{View}_{\cal A} (\enc(m_1), r_{\cal A}),\\
&\qquad\qquad\qquad\qquad\;\;\mathsf{View}_{\cal A}(\enc(m_2), r_{\cal A})) \leq \epsilon
\end{split}
\]
	Here we assume the adversary uses the same random coins $r_{\cal A}$ for the encoding of two messages.

\smallskip
\noindent
ii) {\em Reliability:} For any message $m$ that is encoded to $c$ by the sender, and corrupted to $y = c + e$ by the $(\rho_r, \rho_w)$- AWTP channel, the probability that the receiver outputs the correct information $m$ is at least $1 - \delta$. Receiver will output $\perp$ with probability no more than $\delta$ and will never output an incorrect message.  That is, 
\[
\bP[\dec(\enc(m)+e)=\perp]\leq \delta
\]

\end{definition}
An AWTP code is {\em perfectly secure} if $\epsilon=0$. 



\remove{
Let $\overline{d}(m, m')$ denote  the minimum Hamming distance between any 
pairs of codewords $u\in {\cC}^m$ and $u'\in {\cC}^{m'}$; 
that is,  $\overline{d}(m, m')= \min_{c,c',c\in {\cal C}^m, c'\in {\cal C}^{m'}} d_H(c,c') $.

\begin{lemma}\label{le_dis1}
If an AWTP code is $\delta$-decodable
for  error vectors $e $ of weight $ wt(e)\leq \rho_wN$, for any pair of messages $m, m'\in \cM$, we have  $\overline{d}(m, m')\geq \rho_wN+1$.
\end{lemma}
\begin{IEEEproof}
Let  ${\cal C}^m=\{c\;|\;c=\enc(m, r_{\cal S}), \forall r_{\cal S}\in {\cal U}\}$ denote the set of codewords that are encodings of a
message $m$. For a pair of messages  $m, m'\in {\cal M}$, let  ${\cC}^m$ and ${\cC}^{m'}$ denote the two
subsets of encodings of the two messages, respectively. 

From $\delta$-decodability of  the AWTP-code 
we have $\overline{d}(m, m')\geq  \rho_wN+1$.
\end{IEEEproof}
}

\begin{definition}
For a fixed $\epsilon>0$, an {\em $\epsilon$-secure AWTP code family} is a  family $\mathbb{C}=\{C^N\}_{N\in \mathbb{N}}$ of $(\epsilon, \delta_N)$-AWTP codes  indexed by $N \in \mathbb{N}$, for a $(\rho_r, \rho_w)$-AWTP channel. 
When $\epsilon=0$, the  family is called  a  {\em perfectly secure AWTP code family.} 
\end{definition}

\begin{definition}\label{def_familyrate}
For a  family $\mathbb{C}$ of \edwtp\;codes the {\em rate $R(\mathbb{C})$ is achievable}
if for any $\xi>0$, there exists  $N_0$ such that for any $N\geq N_0$, we have, $\frac{1}{N}\log_{|\Sigma|}|{\cal M}_N|\geq R(\mathbb{C})-\xi$, and the probability of decoding error is $\delta\leq \xi$.
\end{definition}

We use the achievable rate of a code family for an AWTP channel to define secrecy capacity of the channel.
\begin{definition}\label{def_secrecycapacity}
The {\em $\epsilon$-secrecy} ({\em perfect secrecy) capacity of a $(\rho_r, \rho_w)$-AWTP channel denoted by ${\bf C}^\epsilon$} (${\bf C}^0$), is the largest achievable rate of all $(\epsilon,\delta)$-AWTP ($(0,\delta)$-AWTP) code families $\mathbb{C}$ for the channel. 
\end{definition}

The following upper bounds are derived in
 \cite{PS13}.

\begin{lemma}\label{le_up1}\cite{PS13}
The $\epsilon$-secrecy capacity of a $(\rho_r, \rho_w)$-AWTP channel satisfies the
 upper bound,
\[
{\bf C}^{\epsilon}\leq 1-\rho_r-\rho_w+2\epsilon \rho_rN\log_{|\Sigma|} (1+\frac{1}{\epsilon})
\]
The upper bound for the perfect secrecy capacity of a $(\rho_r, \rho_w)$-AWTP channel is,
$
{\bf C}^0\leq 1-\rho_r-\rho_w.
$
\end{lemma}


\section{An  Efficient Capacity Achieving AWTP-Code
}\label{sec_construction}

The general approach to the construction was outlined in Section \ref{sec_intro_codefamily}.  
Below we recall the definition of the building blocks, and give our
instantiations, and construction of the code.


\subsubsection{Algebraic Manipulation Detection Code (AMD code)}\label{sec_amd}
Consider a storage device $\Sigma({\cal G})$ that  holds an element $x$ from a group $\cal G$. The storage $\Sigma({\cal G})$ is private
 but can be manipulated by the adversary by adding  $\Delta\in {\cal G}$.  AMD code allows the manipulation to be detected.

\begin{definition}[AMD-code\cite{CDFP08}]\label{def_amd}
An $({\cal X}, {\cal G}, \delta)$-Algebraic Manipulation Detection code $($$({\cal X}, {\cal G}, \delta)$-AMD code$)$ consists of two algorithms $(\mathsf{AMDenc}, \mathsf{AMDdec})$. 
Encoding given by, $\mathsf{AMDenc}: {\cal X} \rightarrow {\cal G}$,  is probabilistic and   maps an element of a set $\cal X$ to an element of an additive group $\cal G$. 
Decoding,  $\mathsf{AMDdec}: \cal G \rightarrow {\cal X} \cup \{\perp\}$, is deterministic and we have $\mathsf{AMDdec}(\mathsf{AMDenc}(x)) = x$, for any $x \in {\cal X}$.  Security of AMD codes is defined by requiring, 
\begin{eqnarray} \label{amd}
\bP[\mathsf{AMDdec}(\mathsf{AMDenc}(x) + \Delta) \in \{x,\perp \}] \leq \delta, \,\, 
\end{eqnarray}
for all $x\in {\cal X}, \Delta \in {\cal G}.$
\end{definition}

An  AMD code is {\em systematic} 
if  the encoding has the form $\mathsf{AMDenc} : {\cal X}\rightarrow {\cal X} \times {\cal G}_1 \times {\cal G}_2$, $x\rightarrow (x, r, t=f (x, r))$
for some function $f$ and $r \stackrel{\$}\leftarrow {\cal G}_1$. 
The decoding function $\mathsf{AMDdec}(x, r, t) = x$ if and only if $t=f (x, r)$ and $\perp$ otherwise.

We use a systematic AMD-code that is based on the extension of the construction 
 in \cite{CDFP08} to extension fields.
Let $\phi$ be a bijection between vectors $\bf v$ of length $N$  over $\bF_q$,
 and elements in $\bF_{q^N}$, and let $\ell$ be an integer such that $\ell + 2$ is not divisible by $q$. Define the function $\mathsf{AMDenc}: \bF_{q^N}^\ell \rightarrow \bF_{q^N}^\ell \times \bF_{q^N} \times \bF_{q^N}$ by $\mathsf{AMDenc}(x) = (x,r,f(x,r))$ where
\[
f(x, r)=\phi^{-1}\left(\phi(r)^{\ell+2}+\sum_{i=1}^{\ell}\phi(x_i)\phi(r)^i\right)\mod q^N
\]

\begin{lemma}\label{le_amd}
For the AMD-code above, 
given a codeword $(x,r,t)$, the success chance of an adversary 
that  has no information about $(x,r,t)$, in constructing a
new codeword  $(x',r',t')= (x'=x+\Delta x, r'=r+\Delta r, t' =t+\Delta t)$,
that  passes the  verification $t' = f(x', r')$ is 
at most $\frac{\ell+1}{q^N}$. 
\end{lemma}

\subsubsection{Subspace Evasive Sets}\label{sec_ses}


We briefly introduce subspace evasive sets. More details can 
be found in 
\remove{
 The details of construction, the encoding function of subspace evasive sets, and the intersection algorithm of subspace evasive sets and v-dimension space, are in 
}
Appendix \ref{ap_ses}.

\begin{definition}[Subspace Evasive Sets\cite{DL12,Gur11}]
Let $\mathcal{S}\subset \bF_q^n$. We say $\mathcal{S}$ is $(v, \ell_{\mathsf{SE}})$-subspace evasive if for all $v$-dimensional affine subspaces $\mathcal{H}\subset \bF_q^n$, we have $|\mathcal{S}\cap \mathcal{H}|\leq \ell_{\mathsf{SE}}$.
\end{definition}

Dvir {\em et al.} \cite{DL12} show that there is an efficient construction for subspace evasive sets $\cS\subset \bF_q^{n}$, and an efficient {\em intersection algorithm} to compute $\cS \cap \cH$ for  any $v$-dimensional subspace $\cH\subset \bF_q^n$. 


\begin{lemma}\cite{DL12}
Let $v, n_1\in \mathbb{N}$, $w=v^2$, $n=\frac{n_1}{w-v}w$ and $\bF_q$ be a finite field. Then there is a $(v, v^{v\cdot C\log\log v})$-subspace evasive set $\cS \subset \bF_q^n$. For any vector ${\bf v}\in \bF_q^{n_1}$, there is a bijection which maps $\bf v$ into an elements of the subspace evasive set. That is  $$\mathsf{SE}: {\bf v}\rightarrow {\bf v}'\in \cS$$
\end{lemma}

\begin{lemma}\cite{DL12}
Let $\cS\subset \bF_q^n$ be the $(v, \ell_{\mathsf{SE}})$-subspace evasive set. There exists an algorithm that, given a basis for any $\mathcal{H}$, output $\cS\cap \mathcal{H}$ in $\mathcal{O}(v^{v\cdot \log\log v})$ time.
\end{lemma}

\subsubsection{Folded Reed-Solomon Code (FRS code)} \label{FRS_Code)}

A error correcting code $C$ is a subspace of $\bF_q^N$.
The rate of the code is $\log_2 |C|/N$.
A code $C$ of length $N$ and rate $R$ is $(\rho, \ell_{\mathsf{List}})$-list decodable if the number of codewords within distance $\rho N$  from any received word is at most $\ell_{\mathsf{List}}$. 
List decodable codes can potentially correct up to 
$1-R$ fraction of errors, which is twice 
that of unique decoding. This is however at the cost of outputting a list of possible sent codewords (messages).
Construction of  good code with efficient list decoding algorithms is an important research question.
An explicit construction of a list decodable code that achieves the list decoding capacity $\rho=1-R-\varepsilon$ is given by Guruswami et al. \cite{Gur11}. The code is called {\em Folded Reed-Solomon codes (FRS codes)}, defined by Guruswami et al. \cite{Gur11}, gives  an explicit construction for list decodable codes
 that achieve the list decoding capacity $\rho=1-R-\varepsilon$.
The code has polynomial time encoding and decoding algorithms.

\begin{definition}\cite{Gur11}
A $u$-Folded Reed-Solomon code is an error correcting  code with block length $N$ over $\bF_q^u$ and $q>Nu$. The message of an FRS code is written
 in the form of a polynomial $f(x)$ with degree $k$ over $\bF_q$. 
The FRS codeword corresponding to the message is a vector over $\bF^u_q$ where each component is a $u$-tuple $(f(\gamma^{ju}), f(\gamma^{ju+1}),\cdots, f(\gamma^{ju+u-1}))$, $0 \leq j < N$, where $\gamma$ is a generator of $\bF_q^*$, the multiplicative group of $\bF_q$. A codeword of a $u$-folded Reed-Solomon code of length $N$ is in one-to-one correspondence with a codeword $c$ of a Reed-Solomon code of length $uN$, and is obtained by grouping together $u$ consecutive components of $c$.
We use $\mathsf{FRSenc}$ to  denote the encoding algorithm of the FRS code. 
$u$ is called the {\em  folding parameter} of the FRS code. 
\end{definition}

We will use the {\em linear algebraic FRS decoding algorithm}  of these codes \cite{Gur11}\remove{that is outlined in the appendix \ref{decode_FRS}}  (Appendix \ref{decode_FRS}). The following Lemma gives the decoding capability of linear algebraic FRS code.

\begin{lemma} \cite{Gur11} \label{le_fd} For a Folded Reed-Solomon code of block length $N $ and rate $R = \frac{k}{uN}$, the following holds for all integers $1\leq v\leq u$. Given a received word $y \in (\bF_q^{u} )^N$ agreeing with $c$ in at least a fraction,
\[
N-\rho N>N(\frac{1}{v+1}+\frac{v}{v+1}\frac{uR}{u-v+1})
\]
one can compute a matrix ${\bf M} \in \bF_q^{k\times (v-1)}$ and  a vector ${\bf z} \in \bF^k_q$  such that the message polynomials $f \in \bF_q[X]$ in the decoded list are contained in the affine space ${\bf M}{\bf b}+{\bf z}$ for ${\bf b} \in \bF^{v-1}_q$ in $O((Nu\log q)^2)$ time.
\end{lemma}


\subsection{An Explicit Capacity Achieving $(0, \delta)$-AWTP Code Family}\label{sec_awtpconstr}

\remove{We construct an AWTP code family  $\mathbb{C}=\{C^N\}_{N\in \mathbb{N}}$ 
for a $(\rho_r, \rho_w)$-AWTP channel. The $(0, \delta)$-AWTP code $\mathbb{C}$ has }
Let ${\cal M}$ denote the message space, $N$ denote the code length and the encoding and decoding algorithms be,  $\enc_{N}$ and
$\dec_{N}$, respectively.
The message, also referred to as the information block of  the AWTP code, 
is ${\bf m}=\{m_1,\cdots, m_{uRN}\}\in {\cal M}$ where $m_i\in \mathbb{F}_q$. Let $\cS$ be a $(v, v^{C\cdot v\cdot\log\log v})$-subspace evasive set in $\bF_q^n$. 
 Let $u$ and $v$ denote the folding and  the interpolation parameters of the 
FRS code, respectively.
Let  $q$ be a prime number larger than $Nu$, $\gamma$ be a primitive element of $\mathbb{F}_q$,  $\ell=\lceil uR \rceil$, $w=v^2$, $b=\lceil\frac{\ell N+2N}{w-v}\rceil$, $n_1=(w-v)b$, $n=wb$, $\mathsf{SE}: \bF_q^{n_1}\rightarrow \cS$ be the bijection of subspace evasive set.

The construction of encoder and decoder for  $C^N$ is given in Figure \ref{algo_en}.
\vspace{3mm}

\begin{center}
{\bf Figure} \ref{algo_en}
\end{center}
\begin{framed}
\noindent{\bf Encoding}:
For a code rate $R$, the sender $\cal S$ does the following. 

\begin{enumerate}\label{algo_en}
\item Start with the information block $\bf m$ of length $uRN$.  Append sufficient zeros $N(\ell-uR)$
to construct a vector ${\bf x}$ of length $N \ell$; that is,
${\bf x}=\{{\bf m}|| 0,\cdots, 0\}$.

\item  Generate a random vector ${\bf r}$ with length $N$ over $\bF_q$.  Use the AMD construction in section \ref{sec_amd} to construct the AMD codeword $\{{\bf x}, {\bf r}, {\bf t}\}$.  That is,
$\mathsf{AMDenc}({\bf x})=\{{\bf x}, {\bf r}, {\bf t}\}$. The length of AMD code is $\ell N+2N$.

\item  Extend  the AMD codeword to length $n_1$ by appending zeros. Encode AMD code into an element  $\bf s$ of the subspace evasive set $\cal S$. The length of $\bf s$ is $n$. That is 
$${\bf s}=\mathsf{SE}({\bf x}, {\bf r}, {\bf t}||0,\cdots, 0)$$

\item Append a random vector ${\bf a}=\{a_1 \cdots a_{u\rho_rN}\}\in \mathbb{F}_q^{u\rho_rN}$ to $\bf s$ to form a vector that will be  the message of the FRS code, and interpret 
that as coefficients of the polynomial $f(x)$ over $\bF_q$. 
That is $\{f_0,\cdots, f_{k-1}\}=({\bf s}||{\bf a})$.
We have $k=deg(f)+1=  u\rho_rN+ n$. 

\item Use $\mathsf{FRSenc}$ 
to 
construct the  FRS codeword $c=\mathsf{FRSenc}(f(X))=
\{c_1,\cdots ,c_N\}$, and $c_i=\{f(\gamma^{i(u-1)}), \cdots, f(\gamma^{iu-1})\} \in \bF_q^u, \, i=1,\cdots, N$.
\end{enumerate}

\medskip
\noindent{\bf Decoding}: The receiver $\cal R$ does the following:
\begin{enumerate}
\item 
Let  $y=c+e$, and $w_H(e)\leq \rho_wN$.
 The $i$-th component of $y$ is $y_i=\{y_{i, 1}, \cdots, y_{i, u}\}$ for $i=1, \cdots, N$.

\item Use the FRS decoding algorithm  $\mathsf{FRSdec}(y)$ to
output a matrix ${\bf M}\in \mathbb{F}_q^{k\times v}$ and a vector ${\bf z}\in \mathbb{F}_q^k$, such that the 
codewords in the output list 
are, ${\cal L}_{\mathsf{FRS}} = 
 {\bf Mb}+{\bf z}$. 
${\bf M}$ has $k$ rows  each giving a component of the output vector as a linear combination of $\{b_1,\cdots, b_{v}\}$. 
Let   $\cH$ denote  the space which is generated by the first $n$ equations. That is $$\cH={\bf M}_{n\times v}{\bf b}+{\bf z}_n, {\bf b}\in \mathbb{F}_q^{v},$$
where ${\bf M}_{n\times v}$ is the first $n$ rows of the submatrix of ${\bf M}_{n\times v}$ and ${\bf z}_n$ is the first $n$ elements of $\bf z$.

\item The decoder calculates the intersection $\cS\cap \cH$ and outputs a list ${\cal L}$ with size at most $v^{C\cdot v\cdot\log\log v}$. Each ${\bf s}_i\in {\cal L}$ corresponds to an AMD codeword $\{{\bf x}_i, {\bf r}_i, {\bf t}_i\}$. 

\item For each AMD codeword $\{{\bf x}_i, {\bf r}_i, {\bf t}_i\}$, the decoder verifies ${\bf t}_i=f({\bf x}_i, {\bf r}_i)$. If there is a unique valid AMD codeword, the decoder outputs the first $uRN$ components of $\bf x$ as the correct message $\bf m$.
Otherwise,  outputs $\perp$.

\end{enumerate}
\end{framed}
\vspace{3mm}

We prove secrecy and reliability, and derive the rate of AWTP code family.

\begin{lemma}[Secrecy]\label{le_privacy}
The AWTP code $C$ provides perfect security for $(\rho_r, \rho_w)$-AWTP channel. 
\end{lemma}


\begin{IEEEproof}
 We show that an AWTP codeword sent over an $(\rho_r, \rho_w)$-AWTP channel leak no information about 
the encoded subspace evasive sets element  $\bf s$ 
and so 
 the message $\bf m$ will remain perfectly secure. 
 Let $S, A, C^{[r]}$ denote  the random variables corresponding to $\bf s$, ${\bf a}$ and $c^{[r]}=\{c_{j_1},\cdots,c_{j_{\rho_rN}}\}$, respectively.
For an adversary  observation 
$\{c_{i_1},\cdots,c_{i_{\rho_rN}}\}$ with $c_{i_j}=\{c_{i_j, 1},\cdots, c_{i_j, u}\}\in \mathbb{F}_q^u$, 
using the FRS encoding  equations, the adversary has the following  $u\rho_r N$ equations.
\[
\begin{bmatrix}
1& \gamma^{(i_1-1)u}  & \cdots  & \gamma^{(i_1-1)u(k-1)}\\ 
\vdots & \vdots  & \cdots & \vdots \\ 
1 &  \gamma^{i_1u-1}& \cdots  & \gamma^{(i_1u-1)(k-1)}\\ 
\vdots & \vdots  & \cdots & \vdots \\ 
1 & \gamma^{(i_{\rho_rN}-1)u}  & \cdots  & \gamma^{(i_{\rho_rN}-1)u(k-1)} \\ 
\vdots & \vdots & \cdots & \vdots \\ 
1 & \gamma^{i_{\rho_rN}u-1} & \cdots & \gamma^{(i_{\rho_rN}u-1)(k-1)}
\end{bmatrix}\times
\begin{bmatrix}
{\bf s}\\ 
{\bf a}
\end{bmatrix}=
\begin{bmatrix}
c_{i_1, 1}\\ 
\vdots\\ 
c_{i_1, u}\\ 
\vdots\\ 
c_{i_{\rho_rN},1}\\ 
\vdots\\ 
c_{i_{\rho_rN,u}}
\end{bmatrix}
\] 
It is easy to see that  $\bf s$  together with the randomness 
${\bf a}$ uniquely determines ${\bf c}^{[r]}$. This gives,  
\begin{equation}\label{ap_eq_secur15}
\bP(C^{[r]}={\bf c}^{[r]}\;|\; \{S, A\}=\{{\bf s}, {\bf a}\})=1.
\end{equation}
Conversely,
for given values of $\bf s$ and $\{c_{i_1}, \cdots, c_{i_{\rho_rN}}\}$,
and noting that the coefficient matrix is Vandermonde, there exists a unique solution for the ${u\rho_rN}$ unknown components
of
${\bf a}=\{a_1, \cdots, a_{u\rho_rN}\}\in\mathbb{F}_q^{u\rho_rN}$. That is
\begin{equation}\label{ap_eq_secur14}
\bP(A={\bf a}\;|\;\{S, C^{[r]}\}=\{{\bf s}, {\bf c}^{[r]}\})=1
\end{equation}
Since $\bf a$ is chosen  uniformly and independent of $\bf s$, we have 
\begin{equation}\label{ap_eq_secur13}
\bP(A={\bf a}\;|\; S={\bf s})=\frac{1}{q^{u\rho_rN}}
\end{equation}

\noindent From (\ref{ap_eq_secur15}),(\ref{ap_eq_secur14}), and (\ref{ap_eq_secur13}) we have,
\begin{align*}
&\bP(C^{[r]}={\bf c}^{[r]}, A= {\bf a}\;|\; S={\bf s})\\
&\qquad= \bP(A={\bf a}| \{S, C^{[r]}\}=\{{\bf s}, {\bf c}^{[r]} \}) 
\bP(C^{[r]}= c^{[r]}| S={\bf s})\\
& \qquad
= \bP(C^{[r]}={\bf c}^{[r]}| \{S, A\}=\{ {\bf s}, {\bf a} \})
 \bP(A= {\bf a}| S={\bf s}),
\end{align*}
which implies for any $\bf s$, 
\begin{eqnarray}
\bP(C^{[r]}={\bf c}^{[r]}| S={\bf s})=\frac{1}{q^{u\rho_rN}}.
\end{eqnarray}
This means that  for any two elements ${\bf s}_1$ and
 ${\bf s}_2$ of the  subspace evasive sets, 
\[
\begin{split}
&\SD(\mathsf{View}_{\cal A}\;|\;{\bf s}_1, \mathsf{View}_{\cal A}\;|\; {\bf s}_2)\\
&=\sum_{{\bf c}^{[r]} \in \mathsf{View}_{\cal A}}\frac{1}{2}|\bP({\bf c}^{[r]}|{\bf s}_1)-\bP({\bf c}^{[r]}|{\bf s}_2)|=0
\end{split}
\]
\end{IEEEproof}

\begin{lemma}[Reliability]\label{le_decodingerror1}
The failure probability of 
$\dec_N$ is bounded by $\delta_N\leq \frac{v^{C\cdot v\cdot\log\log v}}{q^N}$.
\end{lemma}



\begin{IEEEproof}
The FRS  decoder 
 outputs a list of  elements of the subspace evasive ${\bf s}_i\in {\cal L}$ with list size at most $\ell_{\mathsf{SE}}\leq v^{C\cdot v\cdot \log\log v}$. Each element corresponds to a
unique AMD codeword  $\{{\bf x}_i, {\bf r}_i, {\bf t}_i\}=\mathsf{SE}^{-1}({\bf s}_i)$. 

We first show that the correct message $\bf m$ will be always output by the receiver. Denote the AMD codeword 
corresponding to 
the  message $\bf m$ as $\{{\bf x}, {\bf r}, {\bf t}\}=\mathsf{AMDenc}({\bf m}||0,\cdots,0)$. 
The list decoding algorithm outputs  
codewords that are at distance at most $\rho_wN$ of the received word and so include
the original codeword. The bijection function $\mathsf{SE}$, encodes  the AMD codeword into an element of the subspace evasive set ${\bf s}\in \cS$ that 
 belongs to the decoded list ${\bf s}\in \cH$  
that 
passes AMD verification. That is,
\[
\mathsf{SE}({\bf x}, {\bf r}, {\bf t}||0,\cdots, 0)\in \cL=\cS\cap \cH\;\; \mathsf{and}\;\; {\bf t}=f({\bf x}, {\bf r})
\]

Second, we show  that
the probability that any other codeword in the list 
is a valid AMD codeword is small.
That is we will show that,
\[
\bP(\{{\bf x}', {\bf r}', {\bf t}'\}=\mathsf{SE}^{-1}({\bf s}')\wedge {\bf s}'\in \cL\wedge {\bf t}'=f({\bf x}', {\bf r}'))\leq \frac{\ell}{q^N}
\]
From Lemma \ref{le_privacy}, the adversary has no information about the encoded subspace evasive sets element $\bf s$ and the AMD codeword $\{{\bf x}, {\bf r}, {\bf t}\}=\mathsf{SE}^{-1}({\bf s})$ and so the adversary error, 
$\{\Delta {\bf x}_i={\bf x}'-{\bf x}, \Delta {\bf r}_i={\bf r}'-{\bf r}, \Delta {\bf t}_i={\bf t}'-{\bf t}\}$, is  
 independent of $\{{\bf x}, {\bf r}, {\bf t}\}$. According to Lemma \ref{le_amd}, the probability that the tampered AMD codeword, $\{{\bf x}', {\bf r}', {\bf t}'\}$, passes the verification is no more than $\frac{\ell}{q^N}$. 

Finally, we show the unique correct message output by receiver with probability at least $1-\frac{v^{C'\cdot v\cdot \log\log v}}{q^N}$. 
The list size is at most $v^{C\cdot v\cdot \log\log v}$ and $\ell \leq u=v^2$. So the probability that any $\{{\bf x}', {\bf r}', {\bf t}'\}\neq \{{\bf x}, {\bf r}, {\bf t}\}$ in decoding list 
 pass the verification ${\bf t}'=f({\bf x}', {\bf r}')$, 
 is no more than $\frac{v^{(C+2)\cdot v\cdot \log\log v}}{q^N}$. That is
\[
\begin{split}
&\bP(\bigcup_{ {\bf s}'\in \cL}\{{\bf x}', {\bf r}', {\bf t}'\}=\mathsf{SE}^{-1}({\bf s}')\wedge {\bf t}'=f({\bf x}', {\bf r}'))\\
&\leq \sum_{ {\bf s}'\in \cL}\bP(\{{\bf x}', {\bf r}', {\bf t}'\}=\mathsf{SE}^{-1}({\bf s}')\wedge {\bf t}'=f({\bf x}', {\bf r}'))\\
&\leq \sum_{ {\bf s}'\in \cL}\bP({\bf t}'=f({\bf x}', {\bf r}'))
\leq \frac{\ell|\cL|}{q^N}
\leq \frac{v^{(C+2)\cdot v\cdot \log\log v}}{q^N}
\end{split}
\]

\end{IEEEproof}

We first find the information rate of the code $C^N$, and then find the achievable rate of the code family $\mathbb{C}$.

\begin{lemma}[Rate of $C^N$]\label{le_informationrate1}
The AWTP code $C^N$ described above provides reliability for a $(\rho_r, \rho_w)$-AWTP channel if the following holds:
\begin{equation}\label{eq_awtppf_rate2}
\rho_w<\frac{v}{v+1}-\frac{v}{v+1}\frac{\frac{v}{v-1}(uR+3)+u\rho_r}{u-v+1}.
\end{equation}
\end{lemma}

Proof is in Appendix \ref{ap_le_informationrate1}.

\begin{lemma}[Achievable Rate of  $\mathbb{C}$]\label{the_informationrate2}
The information rate of the $(0, \delta)$-AWTP code family $\mathbb{C}=\{C^{N}\}_{N\in \mathbb{N}}$ for a $(\rho_r, \rho_w)$-AWTP channel is $R(\mathbb{C})=1-\rho_r-\rho_w$. 
\end{lemma}
\begin{IEEEproof}
For a given  small  $\frac{1}{2}>\xi>0$, let code parameters be chosen as, $\xi_1=\frac{\xi}{13}$,
 $v=1/\xi_1$ and $u=1/\xi_1^2$. 
Finally let, $N_0>(1/\xi)^{C/\xi\log\log 1/\xi}$ where $C>0$ is constant. 
From
\[
\begin{split}
&1- R-\rho_r-12\xi_1\leq\\
&\qquad\qquad\qquad \frac{1}{\xi_1+1}-\frac{1}{\xi_1+1}\frac{\frac{1}{1-\xi_1}(R+3\xi_1^2)+\rho_r}{1-\xi_1+\xi_1^2}
\end{split}
\]
the decoding condition (\ref{eq_awtppf_rate2}) of AWTP code
 is satisfied if,
\begin{equation}\label{pf_awtp_rate1}
\rho_w< 1- R-\rho_r-12\xi_1.
\end{equation}

We choose $R=1-\rho_r -\rho_w -12\xi_1$, the decoding condition of AWTP code will be  satisfied. Now since $\xi= 13\xi_1$, for any $N>N_0$, the rate of the AWTP code $C^N$ is 
\[
\begin{split}
\frac{1}{N}\log_{|\Sigma|}|{\cal M}_N|& = R = 1-\rho_r-\rho_w-12\xi_1\\
&>1-\rho_r-\rho_w-\xi=R(\mathbb{C})-\xi
\end{split}
\]
and the probability of decoding error,
\[
\delta_N\leq (1/\xi)^{C/\xi\log\log 1/\xi}q^{-N}\leq Nq^{-N} \leq \xi
\]
So the information rate of AWTP code family $C$ is $R(\mathbb{C})=1-\rho_r-\rho_w$.

\end{IEEEproof}

The computational time for encoding is $\mathcal{O}((N\log q)^2)$. The decoding of FRS code and intersection algorithm of the subspace evasive set is $\mathcal{O}((1/\xi)^{C/\xi\log\log 1/\xi})$. The AMD verification is $\mathcal{O}((1/\xi)^{C/\xi\log\log 1/\xi}(N\log q)^2)$. So the total computational time of decoding is $\mathcal{O}((N\log q)^2)$.

\begin{theorem}
For any small $\xi > 0$, there is $(0,\delta)$-AWTP code $C^N$ of length $N$ over $(\rho_r,\rho_w)$-AWTP
channel such that the information rate is $R(C^N ) = 1-\rho_r-\rho_w-\xi$, the size of alphabet is $|\Sigma| = \mathcal{O}(q^{1/\xi^2})$
and decoding error $\delta < q^{-\mathcal{O}(N)}$. The computational time is $\mathcal{O}((N\log q)^2)$. The AWTP code family $\mathbb{C}=\{C^{N}\}_{N\in \mathbb{N}}$ achieves secrecy capacity $R(\mathbb{C})=1-\rho_r-\rho_w$ for $(\rho_r, \rho_w)$-AWTP channel.
\end{theorem}


\section{Concluding Remarks}\label{sec_conclude}
$(\rorw)$-AWTP extends Wyner wiretap models \cite{ALCP09} to include 
active corruption at physical layer of communication channel.
Although corruption  in our general model is additive, for $S_w\subset S_r$,
it is equivalent to arbitrary replacement of  code components.
We proposed an efficient construction for a capacity achieving code family for
$\rorw$-AWTP channels. The alphabet size for the code is $\bF_q^u$ where for $\delta<\xi$, 
$u=\mathcal{O}(\frac{1}{\xi^2})$ . That is for small failure probability, larger size alphabet must be
used.
Constructing capacity achieving codes over small (fixed) size alphabets remains an open 
problem.

\remove{

We proposed a  model for active adversaries in wiretap channels, derived secrecy capacity and
gave an explicit construction for a family of capacity achieving codes.
The model is a natural extension of Wyner wiretap models when the adversary is 
a powerful active adversary that uses its partial observation of the 
communication channel to introduce adversarial noise in the channel.
Considering interaction over AWTP channels, secret key agreement problem,
as well as  variations in the setting and
adversarial power 
including considering adversarial and probabilistic noise both,  will be 
 interesting 
directions
for future work.
}

\section*{Acknowledgment}

This research is in part supported by Alberta Innovates Technology Future, in the province of Alberta, Canada.



\appendices


\section{Subspace Evasive Sets}\label{ap_ses}

Recently, Guruswami et al. \cite{Gur11} showed that the subspace evasive sets can be used to reduce the list size of list decoding algorithm. Dvir et al. \cite{DL12} gives a explicit and efficient construction of subspace evasive sets. We briefly introduce Dvir et al. \cite{DL12}'s construction of subspace evasive sets. In detail, we give the definition of subspace evasive set, the construction, the encoding function, and the bound of the size of intersection between subspace evasive sets and any $v$-dimensional space.  

\begin{definition}\cite{Gur11}\cite{DL12}
Let $\cS\subset \bF^n$. We say $\cS$ is $(v, \ell_\mathsf{SE})$-subspace evasive sets if for all $v$-dimensional affine subspaces $\cH\subset \bF^n$, there is $|\cS\cap \cH|\leq \ell_\mathsf{SE}$.
\end{definition}

\subsection{Construction of Subspace Evasive Set}

Let $\bF$ be a field and $\overline{\bF}$ be its algebraic closure. A variety in $\overline{\bF}^w$ is the set of common zeros of one or more polynomials. Given $v$ polynomials $f_1, \cdots, f_v\in \overline{\bF}[x_1, \cdots, x_w]$, we denote the variety as
\[
{\bf V}(f_1,\cdots, f_v)=\{{\bf x}\in \overline{\bF}^w\;|\;f_1({\bf x})=\cdots=f_v({\bf x})=0\}
\]
where ${\bf x}=\{x_1, \cdots, x_w\}$.

For a polynomials $f_1, \cdots, f_v\in \bF[x_1, \cdots, x_{w}]$, we define the common solutions in $\bF^w$ as
\[
\begin{split}
{\bf V}_{\bF}(f_1,\cdots, f_v)&={\bf V}(f_1,\cdots, f_v)\cap \bF^w\\
&=\{{\bf x}\in \bF^w\;|\;f_1({\bf x})=\cdots=f_v({\bf x})=0\}
\end{split}
\] 

We say that a $v\times w$ matrix is strongly-regular if all its $r\times r$ minors are regular for all $1\leq r\leq v$. For instance, if $\bF$ is a field with at least $w$ distinct nonzero elements $\gamma_1, \cdots, \gamma_w$, then $A_{i,j}=\gamma_j^i$ is strongly-regular.

\begin{lemma}(Theorem 3.2 \cite{DL12})
Let $v\geq 1, \varepsilon>0$ and $\bF$ be a finite field. Let $w=v/\varepsilon$ and $w$ divides $n$. Let $A$ be a $v\times w$ matrix with coefficients in $\bF$ which is strongly-regular. Let $d_1>\cdots>d_w$ be integers. For $i\in [v]$ let
\[
f_i(x_1,\cdots, x_w)=\sum_{j=1}^wA_{i,j}x_j^{d_j}
\]
and define the subspace evasive sets $\cS\in \bF^n$ to be $(n/w)$ times cartesian product of ${\bf V}_{\bF}(f_1, \cdots, f_v)\subset \bF^w$. That is
\[
\begin{split}
S&={\bf V}_{\bF}(f_1, \cdots, f_v)\times \cdots \times {\bf V}_{\bF}(f_1, \cdots, f_v)\\
&=\{{\bf x}\in \bF^n: f_i(x_{tw+1},\cdots, x_{tw+w})=0,\\
&\qquad \forall 0\leq t<n/w, 1\leq i\leq v\}
\end{split}
\]
Then $\cS$ is $(v, (d_1)^v)$-subspace evasive sets. 

Moreover, if at least $v$ of the degrees $d_1,\cdots, d_w$ are co-prime to $|\bF|-1$, then $|\cS|=|{\bF}|^{(1-\varepsilon)n}$.
\end{lemma} 

The size of list is bounded by $d_1$ and $v$. If we can bound $d_1$ by $v$, the list size can be only bounded by the $v$-dimension subspace $\cH$.

\begin{lemma}(Claim 4.3 \cite{DL12})
There exists a constant $C>0$ such that the following holds: There is a deterministic algoritm that, given integer inputs $v, N$ so that in $\pl(N)$ time there is prime $q$ and $v$ integers $v^{C\log\log v}>d_1>d_2>\cdots>d_v>1$ such that:
\begin{enumerate}
\item For all $i\in [v]$, $\gcd(q-1, d_i)=1$
\item $N<q\leq N\cdot v^{C\log\log v}$
\end{enumerate}
\end{lemma}

Because we only need to choose $w$ integer $d_1>\cdots>d_w$ and $v$ of the integers are co-prime to $q$, the bound of $d_1$ is $d_1\leq \max(w, v^{C\log\log v})$.

\subsection{Encoding Vector as Elements in $\cS$}

We show the encoding map $\mathsf{SE}: {\bf v}\rightarrow {\bf s}$. 
Assuming there is a vector $\bf v$ of length $n_1$ and $(w-v)|n_1$. First we divide the vector into $\frac{n_1}{w-v}$ blocks. Then for each block ${\bf v}_i$ for $i=1,\cdots, \frac{n_1}{w-v}$, we encode into a block ${\bf s}_i$ using bijection $\varphi$. Then we concatenate each block ${\bf s}_i$ for $i=1,\cdots, \frac{n_1}{w-v}$ and generate $\bf s$ in $\cS$. We give the function $\varphi$ in the following.

\begin{lemma}(Claim 4.1)
Assume that at least $v$ of the degree $d_1, \cdots, d_v$ are co-prime to $|\bF|-1$. Then there is an easy to compute bijection $\varphi: \bF^{w-v}\rightarrow {\bf V}_{\bF}\subset \bF^w$. Moreover, there are $w-v$ coordinates in the output of $\varphi$ that can be obtained from the identity mapping $Id: \bF^{w-v}\rightarrow \bF^{w-v}$. 
\end{lemma}

Let $d_{j_1},\cdots, d_{j_v}$ be the degree among $d_1, \cdots, d_w$ co-prime to $|\bF|-1$ and let $J=\{j_1,\cdots,j_v\}$ and $x_{j_i}^{d_{j_i}}=y_{i}$. On the positions $[w]\backslash J$, the map $\varphi$ takes the elements from $\bF^{w-v}$ to $\bF^{[w]\backslash J}$. For the elements on $J$, there is
\[
\sum_{j\in J}A_{i,j}x_j^{d_j}=-\sum_{j\notin J}A_{i,j}x_j^{d_j}
\]
Let $A'$ be the $v\times v$ minor of $A$ given by restricting $A$ to columns in $J$ and $b_i=-\sum_{j\notin J}A_{i,j}x_j^{d_j}$. Then 
\[
A'y=b
\]
and for each $y$, there is unique solution of $x_{j_i}^{d_{j_i}}=y_{i}\mod q$ because $d_{j_i}$ is co-prime to $q-1$.

\subsection{Computing Intesection}

We show how to compute the intersection $\cS \cap \cH$ given $(v, \ell_{\mathsf{SE}})$ subspace evasive sets $\cS$ and $v$-dimension subspace $\cH$. The subspace evasive sets $\cS$ will filter out the elements in $\cH$ and output a set of elements $\cS\cap \cH$ with size no more than $\ell_{\mathsf{SE}}$.

\begin{lemma}(Claim 4.2 \cite{DL12})
Let $\cS\subset \bF^n$ be the $(v, \ell_{\mathsf{SE}})$-subspace evasive sets. There exists an algorithm that, given a basis of $\cH$, output $\cS\cap \cH$ in $\pl((d_1)^v)$ time.
\end{lemma}

Because $\cH$ is $v$-dimensional subspace and $\cH\subset \bF^n$, there exists a set of affine maps $\{\ell_1, \cdots, \ell_n\}$ such that for any elements ${\bf x}=\{x_1, \cdots, x_m\}\in \cH$, there is $x_i=\ell_i(s_1, \cdots, s_v)$.

We show the result by induction of the number of blocks $i=1,\cdots, n/w$. If $i=1$, let $\cH_1:=\{(x_1, \cdots, x_w): (x_1, \cdots, x_n)\in \cH\}$, the dimension of $\cH_1$ is $r_1\leq v$ and $\cH_{x_1, \cdots, x_w}=\{(x_1, \cdots, x_n)\in \cH: (x_1, \cdots, x_w)\}$ such that $\cH=\cup_{(x_1, \cdots, x_w)\in \cH_1}\cH_{x_1, \cdots, x_w}$, and the dimension of $\cH_{x_1, \cdots, x_w}$ is $v-r_1$.
There is 
\[
\begin{split}
&{\bf V}_{\bF}(f_1,\cdots, f_v)\cap \cH_1\\
&=\{(x_1, \cdots, x_w)=(\ell_1(s_1, \cdots, s_v), \cdots, \ell_w(s_1, \cdots, s_v)):\\ 
&\qquad f_1(\ell_1(s_1, \cdots, s_v), \cdots, \ell_w(s_1, \cdots, s_v))=0,\cdots, \\
&\qquad f_v(\ell_1(s_1, \cdots, s_v), \cdots, \ell_w(s_1, \cdots, s_v))=0\}
\end{split}
\]
We can solve the $v$ equations to get $(s_1, \cdots, s_v)$ and then obtain $(x_1, \cdots, x_w)$.
Since $\cH_1\subset \bF^w$,
\[
{\bf V}_{\bF}(f_1,\cdots, f_v)\cap \cH_1={\bf V}(f_1,\cdots, f_v)\cap \cH_1
\] 
By Bezout's theorem, there is $|{\bf V}(f_1, \cdots, f_v)\cap \cH_1|\leq (d_1)^{r_1}$. So there are at most $(d_1)^{r_1}$ solutions for $(x_1, \cdots, x_w)\in \cH_1$. The computational time of solving the equation system follows from powerful algorithms that can solve a system of polynomial equations (over finite fields) in time polynomial in the size of the output, provided that the number of solutions is finite in the algebraic closure (i.e the ‘zero-dimensional’ case). So for $i=1$, the computational time is at most $\pl((d_1)^{r_1})$ and there are $(d_1)^{r_1}$ solutions for $(x_1, \cdots, x_w)$. 

For every fixed of the first $w$ coordinates, we reduce the dimension of $\cH$ by $r_1$ and obtained a new subspace $\cH_2$ on the remaining coordinates. Continuing in the same fashion with $\cH_2$ on the second block we can compute all the solutions in times $\pl((d_1)^{r_1})\cdot \pl((d_1)^{r_2}) \cdots \pl((d_{1})^{r_{n/w}})$, where $r_1+r_2+\cdots +r_{n/w}=v$. So the total running time is $\pl((d_1)^{v})$.


\section{List Decodable Code}

\subsection{Decoding algorithm of FRS code} \label{decode_FRS}

Linear algebraic list decoding \cite{Gur11} has two main steps: interpolation and message finding as outlined below.

\begin{itemize}
\item Find a polynomial, $Q(X, Y_1, \cdots, Y_v)=A_0(X)+A_1(X)Y_1+\cdots+A_v(X)Y_v$, over $\bF_q$ such that $\mbox{deg}(A_i(X)) \leq D$, for $i=1 \cdots v$, and $\mbox{deg}(A_0(X)) \leq D+k-1$, 
  satisfying $Q(\alpha_i, y_{i_1}, y_{i_2},\cdots ,y_{i_v})=0$ for $1\leq i\leq n_0$, where $n_0=(u-v+1)N$.

\item Find all polynomials $f(X) \in \bF_q[X]$ of degree at most $k-1$, with  coefficients $f_0, f_1 \cdots f_{k-1}$,
    that satisfy, $A_0(X)+A_1(X)f(X)+A_2(X)f(\gamma X)+\cdots+A_v(X)f(\gamma^{v-1}X)=0$,
by solving linear equation system.
\end{itemize}

The two above requirements are satisfied if $f \in \bF_q[X]$ is a polynomial of degree at most $k- 1$ whose FRS encoding  agrees with the received word $\bf y$ in at least $t$ components: 
\[
t>N(\frac{1}{v+1}+\frac{v}{v+1}\frac{uR}{u-v+1})
\]

This means we need to find all polynomials $f(X) \in \bF_q[X]$ of degree at most $k-1$, with coefficients $f_0, f_1, \cdots, f_{k-1}$, that satisfy, 
\[
\begin{split}
&A_0(X)+A_1(X)f(X)+A_2(X)f(\gamma X)+\cdots\\
&+A_v(X)f(\gamma^{v-1}X)=0
\end{split}
\]

Let us denote $A_i(X) = \sum_{j=0}^{D+k-1} a_{i,j}X^j$
for $0 \leq i \leq v$. ($a_{i,j} = 0$ when $i \geq 1$ and $j \geq D$). 
Define the polynomials,
\[
\begin{cases}
\begin{split}
&B_0(X)  =  a_{1,0} + a_{2,0}X + a_{3,0}X^2+ \cdots + a_{v,0}X^{v-1}\\
& \ \ \ \ \ \ \vdots\\
&B_{k-1}(X) = a_{1,k-1} + a_{2,k-1}X + a_{3,k-1}X^2+ \cdots \\
&+  a_{v,k-1}X^{v-1}\\
\end{split}
\end{cases}
\]

We examine the condition that the coefficients of $X^i$ of the polynomial $Q(X) = A_0(X) + A_1(X)f(X) + A_2(X)f(\gamma X) + \cdots + A_v(X)f(\gamma^{v-1} X)=0$ equals $0$, for $i=0\cdots k-1$. This is equivalent to the following system of linear equations for $f_0\cdots f_{k-1}$.

\begin{equation}\label{eq_FRS2}
\begin{split}
&\begin{bmatrix}
B_0(\gamma^0) &0  & 0 &\cdots  &0 \\
B_1(\gamma^0) &B_0(\gamma^1)  & 0 &\cdots  &0 \\
B_2(\gamma^0) &B_1(\gamma^1)  & B_0(\gamma^2) &\cdots  &0 \\
\vdots &\vdots  & \vdots & \vdots & \vdots \\
B_{k-1}(\gamma^0) & B_{k-2}(\gamma^{1}) &B_{k-3}(\gamma^2)  &\cdots  & B_0(\gamma^{k-1})
\end{bmatrix}\\
&\times\begin{bmatrix}
f_0\\
f_1\\
f_2\\
\vdots\\
f_{k-1}
\end{bmatrix}=\begin{bmatrix}
-a_{0,0}\\
-a_{0,1}\\
-a_{0,2}\\
\vdots\\
-a_{0,k-1}
\end{bmatrix}
\end{split}
\end{equation}
The rank of the matrix of (Eqs. \ref{eq_FRS2}) is at least $k-v+1$ because there are at most $v-1$ solutions of equation $B_0(X)=0$ so at most $v-1$ of $\gamma^i$ that makes $B_0(\gamma^i)=0$. The dimension of solution space is at most $v-1$ because the rank of matrix of (Eqs. \ref{eq_FRS2}) is at least $k-v+1$. So there are at most $q^{v-1}$ solutions to (Eqs. \ref{eq_FRS2}) and this determines the size of the list which is equal to $q^{v-1}$.


\section{Proof of Lemma \ref{le_informationrate1}}\label{ap_le_informationrate1}
\begin{IEEEproof}
FRS decoding algorithm $\mathsf{FRSdec}$ requires,
\begin{equation}\label{eq_awtppf_rate1}
N-\rho_wN>N(\frac{1}{v+1}+\frac{v}{v+1}\frac{uR_{\mathsf{FRS}}}{u-v+1})
\end{equation}
The dimension of the FRS code is bounded by, 
\begin{equation}\label{eq13927374}
\begin{split}
k=&uR_{\mathsf{FRS}}N=w\lceil \frac{\ell N+2N}{w-v} \rceil+u\rho_rN\\
\leq& \frac{w}{w-v}(uRN+3N)+u\rho_rN.
\end{split}
\end{equation}
The (\ref{eq13927374}) holds because $\ell\leq uR+1$.
So the decoding condition for FRS code (\ref{eq_awtppf_rate1}) holds if,
\[
\begin{split}
N-\rho_wN>N(\frac{1}{v+1}+\frac{v}{v+1}\frac{\frac{w}{w-v}(uR+3)+u\rho_r}{u-v+1})
\end{split}
\]
From $w=v^2$, it is equivalent to,
\[
\rho_w<\frac{v}{v+1}-\frac{v}{v+1}\frac{\frac{v}{v-1}(uR+3)+u\rho_r}{u-v+1}.
\]

\end{IEEEproof}


\ifCLASSOPTIONcaptionsoff
  \newpage
\fi


\begin{thebibliography}{1}


\bibitem{ALCP09} V. Aggarwal, L. Lai, A. Calderbank, and H. Poor, ``Wiretap Channel Type II with an Active Eavesdropper", \emph{ISIT}, pp. 1944--1948, 2009.

\bibitem{BTV12} M. Bellare, S. Tessaro, and A. Vardy, ``Semantic Security for the Wiretap Channel'', \emph{CRYPTO}, pp, 294--311, 2012.


\bibitem{BB11} M. Bloch, and J. Barros, ``Physical-Layer Security: From Information Theory to Security Engineering", \emph{Cambridge Academic Press}, 2011.

\bibitem{BS13} H. Boche, and R. Schaefer, ``Capacity Results and Super-Activation for Wiretap Channels With Active Wiretappers", \emph{IEEE Transactions on Information Forensic and Security}, vol. 8(9) pp. 1482--1496, 2013.


\bibitem{CDFP08} R. Cramer, Y. Dodis, S. Fehr, C. Padr\'{o}, and D. Wichs, ``Detection of Algebraic Manipulation with Applications to Robust Secret Sharing and Fuzzy Extractors", \emph{EUROCRYPT}, pp. 471--488, 2008.

\bibitem{CK78} I. Csiszar and J. K\"{o}rner, ``Broadcast Channels with Confidential Messages", \emph{IEEE Transaction on Information Theory}, vol. 24(3), pp. 339-–348, 1978.


\bibitem{DL12} Z. Dvir, and S. Lovett,	``Subspace evasive sets", \emph{STOC}, pp. 351--358, 2012. 


\bibitem{Gur11} V. Guruswami, ``Linear Algebraic List Decoding of Folded Reed-Solomon Codes", \emph{IEEE Conference
on Computational Complexity}, pp. 77--85, 2011.


\bibitem{LH78} Leung-Yan-Cheong, and E. Hellman, ``The Gaussian wire-tap channel", \emph{IEEE Transactions on Information Theory},  vol. 24(4), pp. 451 -- 456, 1978.

\bibitem{MV11} H. Mahdavifar, and A. Vardy, ``Achieving the Secrecy Capacity of Wiretap Channels Using Polar Codes". \emph{IEEE Transactions on Information Theory}, vol. 57, pp. 6428--6443, 2011.


\bibitem{MBL09} E. MolavianJazi, M. Bloch, and J. N. Laneman, ``Arbitrary Jamming Can Preclude Secure Communication", \emph{47th Annual Allerton Conference on Communication, Control, and Computing}, pp. 1069--1075, 2009.

\bibitem{OW84} L. Ozarow, and A. Wyner, ``Wire-Tap Channel II", \emph{EUROCRYPT}, pp. 33--50, 1984.


\bibitem{PS13} P. Wang and R. Safavi-Naini, ``A Model for Adversarial Wiretap Channel",  CoRR abs/1312.6457(2013).


\bibitem{SW13} R. Safavi-Naini, and P. Wang, ``Codes for Limited View Adversarial Channels", \emph{ISIT}, pp. 266--270, 2013.

\bibitem{W75} A. Wyner, ``The wire-tap channel", \emph{Bell System Technical Journal}, vol. 54, pp. 1355–-1387, 1975.



\end{thebibliography}
\end{document}